\documentclass{epl}
\usepackage{amsfonts,amssymb,epsfig}


\begin{document}

\newcommand{\text}[1]{\mathrm{#1}}
\newcommand{\w}{\omega}
\newcommand{\EWERT}[1]{\left\langle #1 \right\rangle}
\newcommand{\RE}{\text{Re}\,}

\title{Transport in a classical model of an one-dimensional Mott insulator:
Influence of conservation laws}
\shorttitle{Transport in a classical model}
\institute{Institut f\"ur Theorie der Kondensierten
 Materie, Universit\"at Karls\-ru\-he, D-76128 Karls\-ruhe, Germany}
\date{\today}
\author{M. Garst and A. Rosch}
\maketitle

\begin{abstract}
  We study numerically how conservation laws affect the optical
  conductivity $\sigma(\w)$ of a slightly doped one-dimensional Mott
  insulator.  We investigate a regime where the average distance
  between charge excitations is large compared to their thermal de
  Broglie wave length and a classical description is possible. Due to
  conservation laws, the dc-conductivity is infinite and the Drude
  weight $D$ is finite even at finite temperatures.  Our numerical
  results test and confirm exact theoretical predictions for $D$ both
  for integrable and non-integrable models. Small deviations from
  integrability induce slowly decaying modes and, consequently,
  low-frequency peaks in $\sigma(\w)$, which can be described by a
  memory matrix approach.
\end{abstract}

Conservation laws and slowly decaying modes strongly affect transport
properties\cite{forster}.  The reason is that the presence of
conserved quantities (e.g. momentum) can prohibit the complete decay
of an electrical current $J$ (in the absence of external fields). If a
certain fraction $J_c$ of $J$ does not decay, the dc-conductivity is
infinite and the optical conductivity $ \sigma(\w)$ is characterized
by a finite Drude weight $D$ on top of a regular contribution
$\sigma_{\text{reg}}(\w)$ even at {\em finite} temperature $T$, $\RE
\sigma(\w)=2 \pi D\, \delta(\w)+ \RE\sigma_{\text{reg}}(\w)$.

In quasi one-dimensional (1D) systems conservation laws are especially
important.  Even in the presence of Umklapp scattering from a periodic
potential, certain pseudo-momenta are approximately conserved
\cite{roschPRL}. Further, the low-energy properties of interacting
electrons in one dimension are well described by integrable models
like the Luttinger, Sine-Gordon or, equivalently, the massive Thirring
model, which possess an infinite number of conservation laws. While
generic models are not integrable, the integrability of their
low-energy theories implies that their low-frequency, low-$T$ optical
conductivity can be strongly influenced by the presence of {\em
  approximately} conserved quantities\cite{roschPRL,roschPrep}.


Quantitatively, the influence of conserved quantities (CQ) $P_n$ on
the conductivity can be understood in the following way.  Any linear
combination of the $P_n$ is conserved: the $P_n$ span a vector space.
Components of $J$ ``perpendicular'' to this space will decay. The
component ``parallel'' to it, i.e. the projection $J_c$ of $J$ onto
this space, is conserved and will give rise to an infinite
conductivity. What is a physically meaningful definition of
``perpendicular''?  We define an operator $B$ to be perpendicular to
$A$, if $\EWERT{B}=0$ in a (linear-response) experiment where a field
conjugate to $A$ is applied, i.e. $A \perp B \Leftrightarrow
\chi_{AB}=0$, where $\chi_{AB}$ is the usual linear-response
susceptibility. Therefore, the susceptibilities are the natural scalar
product of operators, $\langle A|B \rangle \equiv T \chi_{AB}$. This
definition is widely used in memory functional theory \cite{forster}.

The total weight of the optical conductivity is given by
$\int_{-\infty}^\infty \RE \sigma(\w) d\w=2 \pi (\chi_{JJ}/2)$. The
non-decaying fraction $J_c$ of $J$ will induce a Drude weight,
\begin{eqnarray}\label{suzukiEqual}
D=\frac{\chi_{J_c J_c}^{\ }}{2}=\frac{\chi_{J_c J}^{\ }}{2} = \frac{1}{2} \sum_{\text{all\ CQs}} \chi_{J P_n}^{\ }
 \left( \hat{\chi}_c^{-1}
\right)_{nm} \chi_{P_m J}^{\ }
\end{eqnarray}
where $(\hat{\chi}_c)_{nm}=\chi_{P_n P_m}^{\ }$ is the matrix of
susceptibilities of conserved $P_n$, and  the
projector onto the space of $P_n$ is written as
$\sum_{nm} |P_n\rangle \left(\hat{\chi}_c^{-1}/T \right)_{nm} \langle
P_m |$. In (\ref{suzukiEqual})  the sum extends over {\em all} the
conserved quantities  of the system.
If the summation includes
only a subset of them  one obtains a lower bound for $D$, known
as the Mazur inequality\cite{suzuki}. 
The physical arguments given above have been put on a more
rigorous basis by Suzuki\cite{suzuki} who derived
Eqn.~(\ref{suzukiEqual}) many years ago.


The interest in (\ref{suzukiEqual}) was recently
revived when Castella, Zotos and others \cite{castella,zotos2}
realized that in integrable 1D quantum models the
Drude weight $D$ is finite at finite $T$, induced by the
conserved quantities which prohibit the decay of the current.  In a
number of ambitious numerical studies of (small) quantum systems
\cite{zotosnumerical,herveMobility,kirchner}, the surprising
observation of a finite Drude weight at $T>0$ in systems like the
Hubbard model was qualitatively confirmed. Moreover, Castella and
Zotos\cite{herveMobility} found that the mobility of a single particle
coupled to Fermions is enhanced close to an integrable point.
However, a number of fundamental issues remained to be answered,
partly due to numerical problems and a lack of a theory to investigate
the consequences of small deviations from integrability.

The goal of the study reported here is threefold: first, we want to
test the important exact expression (\ref{suzukiEqual}) for $D$
numerically for a simple system.  Previous results have cast serious
doubts on its validity \cite{zotosnumerical}.  Our second objective is
to investigate how {\em weakly violated} conservation laws influence
the conductivity\cite{roschPRL,roschPrep} and to test analytical
approximations for $\sigma(\w)$.  A third goal is to calculate the
optical conductivity close to a Mott transition in a 1D
system. This is directly relevant to recent experiments in Bechgaard
salts\cite{bechgaards} where well defined low-frequency peaks have
been observed in $\sigma(\w)$ below a Mott gap.

The derivation of an effective model of charge excitations close to a
Mott transition starts from the Luttinger liquid description of a
nearly half-filled band in the presence of Umklapp scattering. Using a
Luther-Emery transformation and a subsequent refermionization (see
e.g. \cite{giamarchi} for details) the charge-sector maps onto a
two-band model of Fermions with a band gap $2 \Delta$, i.e.  a
``semiconductor''. The interactions of the ``particles'' and ``holes''
are repulsive for $1/2 < K <1$ where $K$ parameterizes the strength of
interactions in the Luttinger liquid.  Any finite repulsive
interaction in 1D leads to a total reflection of particles in the
low-density, low-energy limit.  Therefore, Sachdev and Damle
\cite{kedar} argued (for a similar problem of a gapped spin-chain)
that the interaction is well characterized by a hard-core model.
Exactly at half filling, the densitiy of charge excitations is
exponentially small for $T<\Delta$. Their distance is therefore
exponentially large, much larger than their de Broglie wavelength $
h/\sqrt{2 \pi m_\alpha T}$ and for  sufficiently small doping and
$T<\Delta$ a classical description is  possible\cite{kedar}.
A simulation of the resulting classical models allows to test
(\ref{suzukiEqual}) and to obtain the optical conductivity with
high precision and small numerical effort.

We consider a model of two types of particles
with the  kinetic energy,
\begin{eqnarray}
H_{\text{kin}}=\sum_{i=1}^{N_+} \epsilon_+(p_{+, i}) +\sum_{j=1}^{N_-}
 \epsilon_-(p_{-, j}) \quad  \text{where}  \quad
\epsilon_{\alpha}(p)=\sqrt{m_\alpha^2+p^2}
\end{eqnarray}
$N_\pm$  is the number of particles with
charge $\pm1$ and momentum $p_{\pm,i}$. 
$H_{\text{kin}}$ together with the local hard-core repulsion define
our model.
The particles are distributed randomly on a finite ring
 and the initial momenta are assigned
according to the Boltzmann distribution.

The  conductivity is computed from the  ensemble
average of the current-current correlation function
$S(t-t') = \frac{1}{L} \langle J(t) J(t') \rangle$ via the
fluctuation-dissipation theorem (in the classical limit)
\begin{eqnarray}
\RE\sigma(\omega) = 
\frac{1}{2 T} S(\omega) = 2 \pi D \delta(\w)+\RE \sigma_{\text{reg}}(\w) 
\quad \text{with}\quad D=\frac{1}{2 T} \lim_{t\to \infty} S(t).
\end{eqnarray}
$L$ is the size of the system and the current is given by
\begin{eqnarray}
J(t)=\sum_{i=1}^{N_+} v_{+, i}(t) -\sum_{j=1}^{N_-} v_{-, j}(t)
\end{eqnarray}
where $v_{\pm,i} = d \epsilon_\pm(p_{\pm, i})/d p_{\pm,i}$ is
the velocity of the $i$th particle.
Due to  conservation
laws, see below, the dynamics is not ergodic. Therefore, we average over
several thousand runs with different initial conditions. In each run,
each of the $500$ to $1800$ particles undergoes typically 1000 collisions.


\begin{figure}[t]
  \centering
\epsfig{width=.47 \linewidth,file=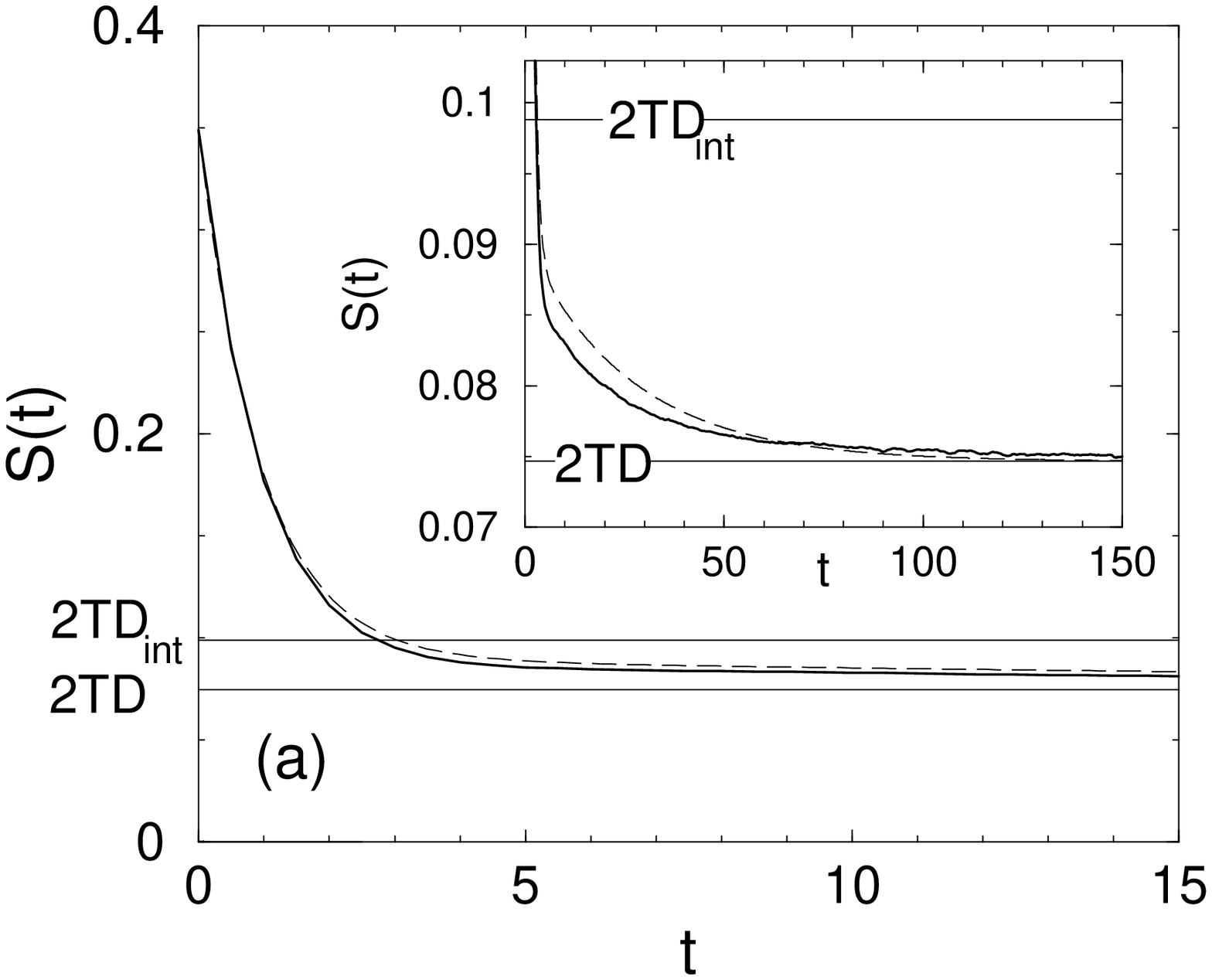}\hspace{0.2cm}
\epsfig{width=.46 \linewidth,file=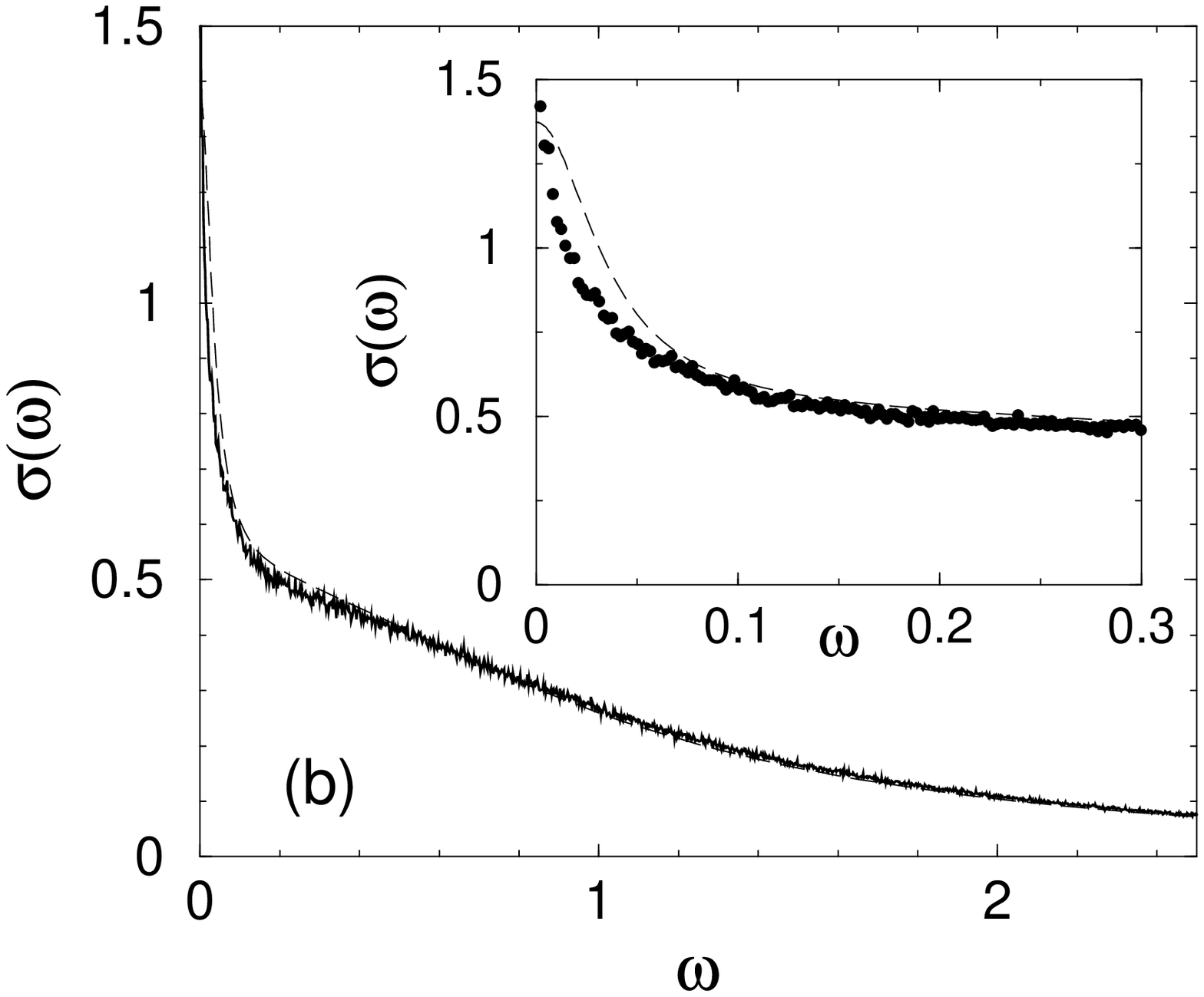}


\caption[]{(a) Current-current 
  correlation function $S(t)$ for $T=0.5$, $N_+/N_-=200/600$ and
  $m_{\pm}=1\pm 0.3$.  In a few collisions $S(t)$ decays down to $2 T
  D_{\text{int}}$ (\ref{DrudeInt}). Due to the approximate
  conservation of $\bar{V}$, it takes much longer (inset) to relax
  down to $2 T D$ (\ref{DrudeNonInt}).  Time and distances are
  measured in units of the average particle spacing $L/(N_++N_-)$; in
  this units the typical time between collisions is $\sim \sqrt{m/T}$
  for low $T$.  A comparison with the analytical approximation
  (\ref{MemMatrixEqu}) and (\ref{memAna}) (dashed line) confirms our
  interpretation that despite the huge mass difference, the proximity
  to the integrable point is at the origin of the slowly decaying
  mode. To allow for a better comparison of the numerical data to the
  analytical approximation, we have multiplied all scattering rates in
  the latter by a factor of $\approx 2$ to fit the short time scale
  ((\ref{memAna}) is only valid up to prefactors of order 1). (b)
  Optical conductivity $\RE \, \sigma(\w)$ (same parameters, doubled
  system size).  Weakly violated conservation laws induce a low
  frequency peak (see inset). The Drude peak $2 \pi D \delta(\w)$ due
  to momentum conservation (\ref{DrudeNonInt}) is not shown.  }
 \label{figSt}\label{figSigma}
\end{figure}



In the following, we compare our numerical results for $D$ with the
 theoretical prediction (\ref{suzukiEqual}). We first consider the
 situation $m_+ \neq m_-$, where we expect that only momentum,
 $P=\sum_1^{N_+} p_{+,i}+\sum_1^{N_-} p_{-,i}$, energy $H$ and 
 charges $N_{\pm}$ are conserved. $H$ and
 $N_{\pm}$ do not contribute to (\ref{suzukiEqual}) as
 $\chi_{HJ}=\chi_{N_{\pm} J}=0$ due to symmetry and therefore
 \begin{eqnarray}\label{DrudeNonInt} D &=& \frac{1}{2}
 \frac{\chi^2_{JP}}{\chi_{PP}} = \frac{1}{2 T L} \frac{\langle J P
 \rangle^2}{\langle P P\rangle}= \frac{T (\Delta N)^2}{2  L}
 \frac{1}{N_+ \EWERT{p^2}_{+}+N_- \EWERT{p^2}_{-}} 
\end{eqnarray}
where $N=N_++N_-$ and $\Delta
N=N_+-N_-$ are the total number of particles and the total charge,
respectively.  Conveniently, in our model all static susceptibilities
are {\em exactly} given by their value in the non-interacting theory 
as we
are considering only local interactions in a classical system, e.g.
$\chi_{J P}=\frac{1}{T L} \left(N_+ \EWERT{p v_+}_{+} - N_- \EWERT{p
v_-}_{-}\right)$, where the {\em
one-particle} expectation value $\EWERT{...}_{\pm}$ in the absence of interactions 
is given by $\EWERT{A}_{\pm}=\int d p A(p) e^{-
\epsilon_{\pm}(p)/T}/\int d p e^{-\epsilon_{\pm}(p)/T} $.


The situation is qualitatively different when the masses of the
``particles'' and ``holes'' are equal, $m_+ = m_-$. The system is
integrable and more conserved quantities exist. Jepsen \cite{jepsen}
calculated the properties of this system ana\-ly\-ti\-cally and we used
 his results \cite{jepsen,kedar} for $\sigma(\w)$ to check our numerics in the
non-relativistic limit $T\ll m$.
The whole momentum distribution of the particles is
conserved for $m_+=m_-$
since in any collision of particles with {\em equal} masses
the momentum is just interchanged.  Hence, all moments
\begin{eqnarray}\label{Pn}
P_n = \sum_{i=1}^{N_+} p_{+, i}^n +\sum_{j=1}^{N_-} p^n_{-, j}
\end{eqnarray}
of the momentum distribution
are conserved.
What is the projection of $J$ onto the space of CQs?  Consider the total
velocity $V=\sum_1^{N_+} v_+ + \sum_1^{N_-} v_- $ which is conserved
for $m_+=m_-$ and differs from $J$ by just a minus sign.  
Since for equal masses 
 $\chi_{J P_n}=\frac{N_+-N_-}{T L} \EWERT{v p^n}_{\pm}=\frac{\Delta N}{N}\chi_{V P_n}$, the projection $J_c$ of $J$ onto the
space of conservation laws $P_n$  is {\em parallel} to 
 $V$ and
\begin{eqnarray}\label{DrudeInt}
D_{\text{int}} =\frac{1}{2} \frac{\chi_{J V}^2}{\chi_{VV}}= 
\frac{1}{2 T  L} \frac{\langle J V
 \rangle^2}{\langle VV\rangle}.
\end{eqnarray}
For $m_+=m_-$, this evaluates to  $\frac{1}{2 T L} 
\frac{(\Delta N)^2}{N} \EWERT{v^2}_{\pm}$.

\begin{figure}[t]
  \centering
\epsfig{width=.48 \linewidth,file=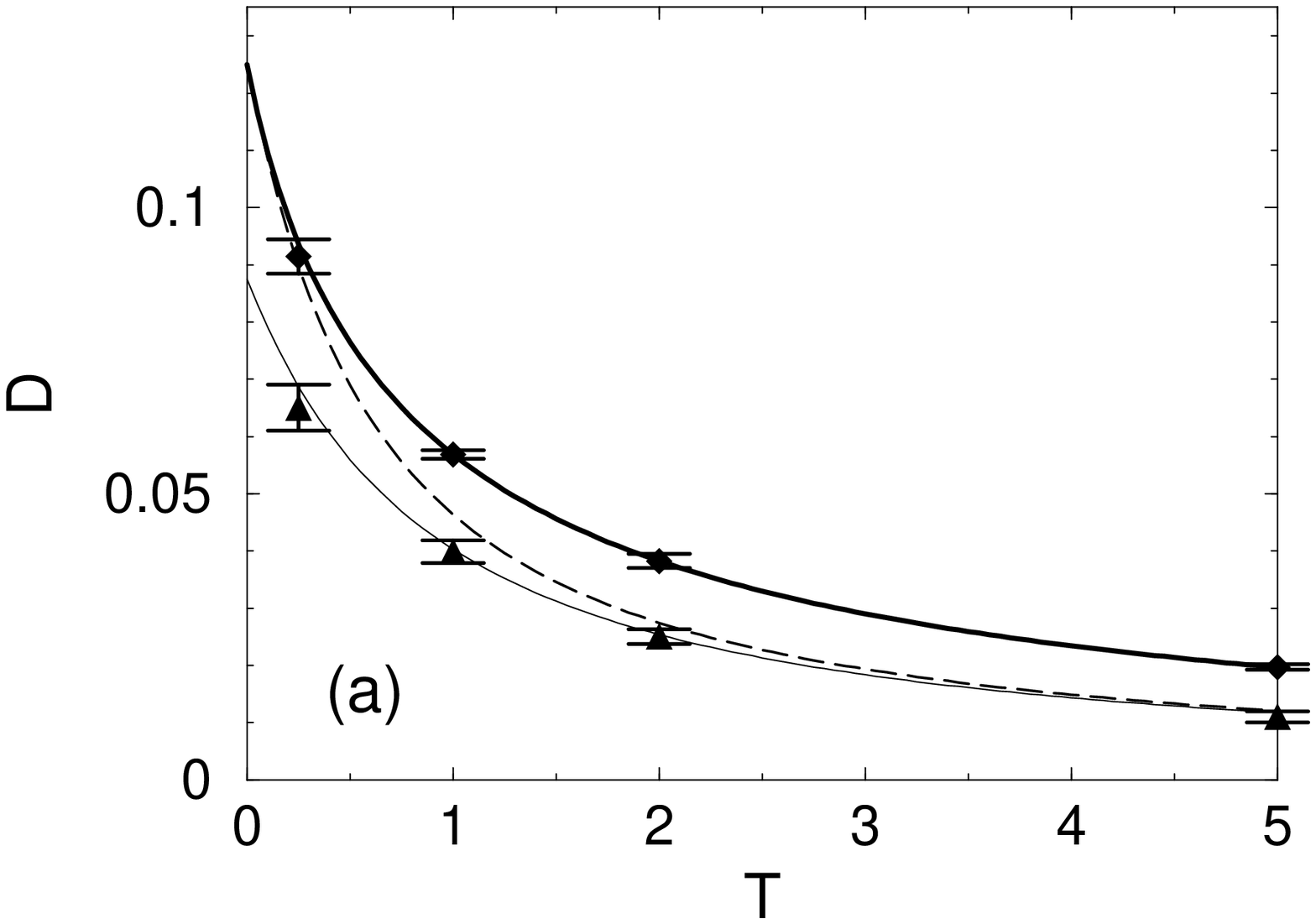}\hspace{.2cm}
\epsfig{width=.47 \linewidth,file=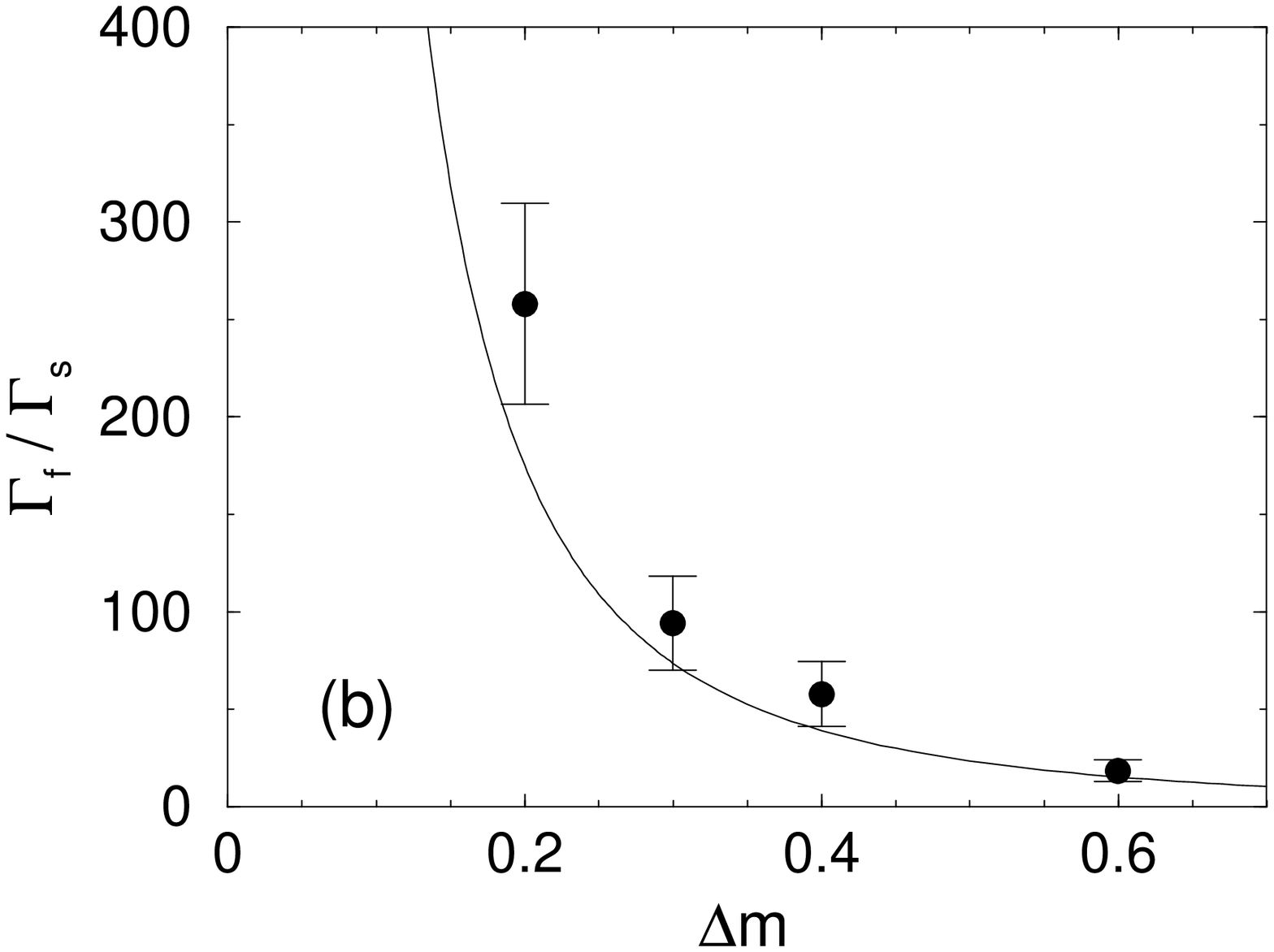}
\caption[]{(a) The numerically determined 
  Drude weights $D(T)$ both for an integrable model with $m_+=m_-=1$,
  (diamonds) and a non-integrable model ($m_+=1$, $m_-=2.7$)
  (triangles) agree within the statistical error bars with the
  theoretical prediction (\ref{DrudeInt}) and (\ref{DrudeNonInt})
  (thick and thin solid lines) ($N_+=3 N_-$).  The dashed line shows
  the Drude weight (\ref{DrudeNonInt}) of a non-integrable system in the limit
$|\Delta m|\to 0^+$, $m_{\pm}\approx 1$.  (b) Ratio of
  the fast and slow decay rate $\Gamma_f/\Gamma_s$ as a function of
  $\Delta m$ for $m_\pm=1\pm \Delta m/2$, $N_+/N_-=100/300$ and
  $T=0.25$.  Close to the integrable point, the slow
  decay rate vanishes $\propto ( \Delta m)^2$.  The points are
  obtained from a fit to $S(t)$ determined from numerical simulations.
  The analytical result (\ref{memAna},\ref{gammass})
  (line) describes the observed behavior qualitatively correct. The
  good agreement at high $\Delta m$ is accidentially, (\ref{memAna})
  is expected to be valid only for small $\Delta m$ and only up to
  prefactors of order $1$.}
\label{figDrude}
\label{figRates}
\end{figure}
In Fig.~\ref{figDrude} we compare the analytical formulas
(\ref{DrudeNonInt}) and (\ref{DrudeInt}) to our numerical results.
Within the statistical errors the computer experiments  clearly confirm the
theory. 
The conservation laws for equal masses $m_+ = m_-$ are very different
from those for $m_+ \neq m_-$. Therefore, $D$ is not a continuous
function of the mass difference $
D(\Delta m \to 0) \neq D_{\text{int}}=D(\Delta m=0)$
as is shown  in Fig.~\ref{figDrude}a (dashed line). 

What will happen if $\Delta m$ is small? This is a physically relevant
situation, e.g. for a Mott insulator with a small gap $\Delta \ll
\epsilon_F$, where corrections which break the particle-hole symmetry
are small but finite. Due to this small mass difference $\Delta m$,
the higher moments $P_n$ in (\ref{Pn}) with $n>1$ will slowly decay
and the following physical picture emerges for $|\Delta m| \ll
m=(m_++m_-)/2$: we decompose the current $J=J_f+J_s$ (as before) into
a ``fast'' component $J_f$ perpendicular to the currents $P_n$ and the
projection $J_s$ of $J$ onto the space of slow variables $P_n$. After
a few collisions, i.e. after the time $t_f$, $J_f$ has decayed
completely and $S(t)=\EWERT{J(t) J(0)}\approx 2 T D_{\text{int}}$ for
$t_f \ll t \ll t_s$ as is shown in Fig.~\ref{figSt}a.  $J_s$ is
exactly conserved for $\Delta m=0$. For finite $\Delta m$ the
projection $J_c$ of $J_s=J_c+J_{sd}$ onto momentum $P=P_1$ is still
conserved but the component $J_{sd}$ perpendicular to $P$ decays on
the scale $t_s \propto 1/(\Delta m)^2$. Accordingly, $S(t)$ decays
further to $S(t\gg t_s) = 2 T D$, where the decay stops
(Fig.~\ref{figSt}a).

The long-time behavior of $S(t)$ determines the low-frequency behavior
of the optical conductivity (Fig.~\ref{figSigma}b): at
zero frequency, $\RE \sigma(\w)$ is characterized by a $\delta$-peak
with weight $2 \pi D$. For small $\w$, we expect a sharp peak of
width $\Gamma_s = 1/t_s$, weight $2 \pi (D_{\text{int}}-D)$ and with a height
approximately given by $(D_{\text{int}}-D)/\Gamma_s$. The {\em slow}
decay rate determines $\RE \,\sigma(\w\to 0)$
\cite{roschPRL}.

In the following, we try to give a crude analytical estimate of the
slow decay rate $\Gamma_s$ within the framework of the (classical)
memory matrix formalism \cite{forster}. 
The correlation function
\begin{eqnarray}
C_{mn}(t) = \langle j_m(t) j_n(0) \rangle
\end{eqnarray}
of a set of currents $j_0,j_1,...,j_N$
is calculated from 
\begin{eqnarray}\label{MemMatrixEqu}
(\hat{M}(\w)\hat{C}^{-1} - i \w) \hat{C}(\w) = \hat{C} \quad \text{with} 
\quad M_{mn}(\w) = \langle \partial_t j_m \mid Q 
\frac{1}{\w-QLQ} Q \mid  \partial_t j_n \rangle
\end{eqnarray}
where $C_{mn} = C(t=0)_{mn}=T \chi_{j_m j_n}$ is the matrix of
susceptibilities, $\hat{M}$ the so-called  memory matrix and $L$ the
Liouville operator defined by the Poisson brackets with the
Hamiltonian $i L A = \{A,H\}$.  The time evolution of
$\hat{M}$ is determined by the projection of $L$ onto the space
perpendicular to the currents $j_k$ with the projection operator $Q =
1 - \sum_{kl} | j_k \rangle (\hat{C}^{-1})_{kl} \langle j_l |$.

The main idea of the memory matrix formalism is that approximations
for $\hat{M}$ are much less ``dangerous'' than approximations for e.g.
$\hat{C}(\w)$, at least if {\em all} relevant slow variables are
included in the space of the $j_n$. In this case, the dynamics in the
perpendicular space defined by $Q$ is dominated by fast processes and
therefore $\hat{M}$ is -- hopefully -- non-singular in the
low-frequency limit. Here, we test the applicability of this approach
in a situation governed by a clear separation of time scales. It is
easy to check \cite{roschPRL} that indeed the exact Drude weight
(\ref{suzukiEqual}) is reproduced if all relevant conservation laws
are included, e.g.  $j_0=J$, $j_n=P_n$, $n=1...\infty$.  Actually, it
is sufficient to keep track of $J$ and its projections onto the $P_n$
and $P$. For $\Delta m=0$, the projection $J_s$ of $J$ onto the space
of the $P_n$ is parallel to $\bar{V}$ with
\begin{eqnarray}
\bar{V} = \sum_{i=1}^{N_+ + N_-} \frac{p_i}{\sqrt{m^2+p_i^2}}.
\end{eqnarray} 
For small $\Delta m$ we replace $m$ by $(m_++m_-)/2$.  We can
therefore restrict our analysis to the three-dimensional space of
operators $j_0=J, j_1=P$ and $j_2=\bar{V}$. One can check that this
approximation is not only valid for $|\Delta m| \ll m$ at any
temperature but also for $T \ll m \pm \Delta m$ for arbitrary $\Delta
m$ because $J_s$ is a linear combination of $P$ and $\bar{V}$ in the
latter case.

While the weights of all features in the optical conductivity can be
calculated exactly, it is much more difficult to determine the decay
rates for small but finite $\Delta m$. To estimate the various
elements of the memory matrix, we use two approximations: we neglect
all correlations between collisions and we neglect the projection
operator $Q$ in (\ref{MemMatrixEqu}):
\begin{eqnarray}
 M_{nm}(t) &\approx& \EWERT{ \partial_t j_n(t) \partial_t j_m(0) }
\approx \sum_i \EWERT{ \Delta j_n^{(i)} \Delta j_m^{(i)} \delta(t_i)}
\delta(t)
 \nonumber \\
&\approx& \frac{N_+ N_-}{L}\EWERT{ \Delta j_n
\Delta j_m |v_+-v_-|}_{p_+ p_-} \delta(t)
\label{memAna}
\end{eqnarray}
where $\Delta j_n^{(i)}(p_+,p_-)$ is the change of $j_n$ at the $i$th
collision at time $t_i$, $\partial_t j_n=\sum_i \Delta j_n^{(i)}
\delta(t-t_i)$ and $\EWERT{ f_{p_+,p_-}}_{p_+ p_-}= \int f_{p_,+p_-}
e^{-(\epsilon_+(p_+)+\epsilon_-(p_-))/T} /\int
e^{-(\epsilon_+(p_+)+\epsilon_-(p_-))/T}$ is the thermal average
over the momenta $p_\pm$ of the two incoming particles with velocities
$v_\pm$ and charge $+$ and $-$, respectively.  Note that collisions of
particles with equal charge do not change $j_n$.  Both approximations
in (\ref{memAna}) are uncontrolled\cite{remark} and will induce errors
of order $1$ in the prefactors of the collision rates. Nevertheless, a
comparison with our numerical data shows that the qualitative
dependence of $\sigma$ as a function of $T$, $\Delta m$ and $\Delta N/N$
is very well described by (\ref{memAna}). For $N_+=N_-$ and $\Delta
m=0$ and $T\to 0$ we obtain e.g. $\sigma(T)=\sqrt{\pi}/(16 \sqrt{ m
  T})$ to be compared to the exact result \cite{kedar,jepsen}
$\sigma(T)=1/\sqrt{2 \pi m T}$, the error is smaller at higher $T$.


The size of $\Delta j_n$ determines the various collision rates.
 $\Delta j_0 \approx 2 (v(p_+)-v(p_-))$ is finite for  $\Delta
m \to 0$ and $M_{00}$ describes the initial fast decay of $J$.
$\Delta j_1$ vanishes due to momentum conservation. Accordingly, the
projection of $J$ onto $P$ will not decay.  $\Delta j_2$ is {\em linear}
in $\Delta m$ as $\bar{V}$ is conserved for $\Delta m=0$ which leads
to a {\em slow} decay rate proportional to $(\Delta m)^2$
\begin{eqnarray}
\Gamma_s/\Gamma_f \propto (\Delta m)^2.
\end{eqnarray}
A comparison of the analytical result with the numerical data shows
that our analytical calculation within the memory matrix formalism
does not only reproduce the exact weights (Fig.~\ref{figDrude}b) but
also describes the relevant decay rates $\Gamma_s$ and $\Gamma_f$
qualitatively and semi-quantitatively correctly. In
Fig.~\ref{figRates}b, we compare the exact decay rates determined from
a fit to the numerically calculated $ S(t)\approx c_1 e^{-\Gamma_f
  t}+c_2 e^{-\Gamma_s t}+c_3$ to the corresponding rates derived from
our analytical approximations (\ref{MemMatrixEqu}) and (\ref{memAna}).
In the limit $\Delta m \to 0$ we obtain
\begin{eqnarray}\label{gammass}
\Gamma_s&=&\frac{M_{22}-M_{02}^2/M_{00}}{
C_{22}-C_{12}^2/C_{11}}, \quad \quad 
\Gamma_f=\frac{M_{00}}{
C_{00}(1-(\Delta N/N)^2)}\label{gammaff}
\end{eqnarray}
where the memory functions $M_{nm}$ are evaluated at $\w=0$. 
At low $T$ we get e.g.
$\Gamma_f \approx 8 \sqrt{T/\pi}$ and
$\Gamma_s\approx 6 (\Delta m)^2 (1-(\Delta N/N)^2) \sqrt{T/\pi}$.

Within our (crude) approximation scheme, $S(t)$ decays {\em exponentially}
to a constant for long times. This is certainly not correct, as e.g.
for $\Delta m=0$, $S(t)\sim const.-1/t^\alpha$ with $\alpha=3$ as
is known from the exact result \cite{jepsen,longTimeTail}. Our
numerical results suggest that the (algebraic) long-time tails are
more pronounced for $\Delta m \neq 0$ but we were not able to extract
the exponent $\alpha$ reliably due to large finite size effects (due
to the hard-core interaction the particles cannot pass each other and
their diffusion is stopped after a distance of order $\sqrt{N}$). The
main physics discussed in this paper, i.e. the existence of a
time-scale $1/\Gamma_s\propto 1/(\Delta m)^2$, seems not to be affected
by the presence of algebraic long-time tails.

How small has $\Delta m$ to be that $\sigma(\w)$ displays a well
defined peak of width $\Gamma_s$ due to the slowly
decaying component of $J$? To decide this question, we compare
$\RE \sigma(\w \to 0^+,\Delta m \to 0^+) \approx 2 \Delta D_s/\Gamma_s$ 
with $\RE \sigma(\w \to 0^+,\Delta m=0)\approx
2 \Delta D_f/\Gamma_f$ where 
$\Delta D_s=D_{\text{int}}-D$ and $\Delta D_f=
\chi_{JJ}/2- D_{\text{int}}$ are  the weights of the slowly and fast 
decaying modes, respectively.
The low frequency peak is visible if
\begin{eqnarray}
1 \gg  \frac{\Delta D_f}{\Gamma_f} / \frac{\Delta D_s}{\Gamma_s}
\approx \frac{(\Delta m)^2 
(1-(\Delta N/N)^2)^2}{2 T^2 (\Delta N/N)^2} 
\end{eqnarray}
where the last equation is valid for low $T$.  The peak is not
observable for $T \lesssim |\Delta m| $ which is a consequence of the
fact that the model is Galileian invariant for $T \to 0$. The slowly
decaying modes are much less relevant for $\Delta N=0$, as $\chi_{P J}
= \Delta N/L$ and no well pronounced low-frequency peak develops for
low $T$ \cite{roschPrep}.

In this paper we have investigated numerically and analytically the
optical conductivity for a model of charge excitations above a Mott
gap. Due to conservation laws, the Drude weight is finite at $T>0$ for
$\Delta N \neq 0$.  We can confirm predictions
(\ref{DrudeInt},\ref{DrudeNonInt}) for the Drude weight and
$\sigma(\w)$ both for an integrable model and in a situation where
conservation laws are weakly broken and low-frequency peaks arise in
$\sigma(\w)$.  One of the surprises of this study was that even for a
really large mass difference, e.g.  $m_{\pm}=1\pm 0.3$, the proximity
of the Hamiltonian to some integrable point with $m_+=m_-$ strongly
influences the low-frequency conductivity (see Fig.~\ref{figSigma}b)).
At least for the model studied here, the memory matrix formalism is a
reliable tool to study this type of physics qualitatively.
Furthermore, we expect a {\em quantitative} agreement in
situations\cite{roschPRL} where it is possible to calculate the memory
matrix perturbatively.  One may speculate that the low-frequency peaks
in the optical conductivity of the Bechgaard salts
are related to approximate conservation laws 
 as they are characterized by tiny weights, unusual
line-shapes and a very long mean-free path\cite{bechgaards}.

We would like to thank N. Andrei, K. Damle, F.~Evers, P.~W\"olfle and X.~Zotos
for helpful discussions and the Emmy-Noether program of the DFG for
financial support.


\begin{thebibliography}{0}
\bibitem{remark} The problem arises
because the hard-core collisions define a strong coupling problem. In
a weak coupling situation\cite{roschPRL} the projection $Q$ can be
neglected in lowest order.
\end{thebibliography}
\end{document}